\documentclass[%
twocolumn,
superscriptaddress,
 amsmath,amssymb,
 aps,
prb,
]{revtex4-2}

\usepackage{graphicx}
\usepackage{dcolumn}
\usepackage{bm}
\usepackage{gensymb}
\usepackage{hyperref}

\usepackage{xcolor}

\usepackage{ulem}
\usepackage{placeins}
\DeclareMathOperator{\erfc}{erfc}
\begin{document}

\title{Spatial structure of surface plasmon resonance under narrow-beam illumination and its implications for sensing}

\author{Tristan Miralles}
\affiliation{Université Clermont Auvergne, CNRS, Clermont Auvergne INP, Institut Pascal, F-63000 Clermont-Ferrand, France}

\author{Rémi Pollès}
\affiliation{Université Clermont Auvergne, CNRS, Clermont Auvergne INP, Institut Pascal, F-63000 Clermont-Ferrand, France}
\affiliation{Institut Universitaire de Technologie de Clermont-Ferrand, Université Clermont Auvergne, F-63000 Clermont-Ferrand, France}

\author{Pauline Bennet}
\affiliation{Aix-Marseille Université, Université de Toulon, CNRS, IM2NP, F-13397 Marseille, France}

\author{Denis Langevin}
\affiliation{Aix-Marseille Université, Université de Toulon, CNRS, IM2NP, F-13397 Marseille, France}

\author{Martine Mihailovic}
\affiliation{Université Clermont Auvergne, CNRS, Clermont Auvergne INP, Institut Pascal, F-63000 Clermont-Ferrand, France}
\author{Emmanuel Centeno}
\affiliation{Université Clermont Auvergne, CNRS, Clermont Auvergne INP, Institut Pascal, F-63000 Clermont-Ferrand, France}

\author{Jack Mock}
\affiliation{Department of Electrical and Computer Engineering and the Center for Metamaterials and Integrated Plasmonics, Duke University, Durham, NC 27708, USA}

\author{David R. Smith}
\affiliation{Department of Electrical and Computer Engineering and the Center for Metamaterials and Integrated Plasmonics, Duke University, Durham, NC 27708, USA}

\author{Antoine Moreau}
\affiliation{Université Clermont Auvergne, CNRS, Clermont Auvergne INP, Institut Pascal, F-63000 Clermont-Ferrand, France}

\date{\today}

\begin{abstract}
The reflection of light on a surface plasmon resonance structure can be understood as the interference between the direct Fresnel reflection and the light radiated back by the plasmon as it propagates along the metal film. When the beam waist is only an order of magnitude larger than the surface plasmon propagation length,  these two contributions separate spatially, and their destructive interference localizes into a dark fringe within the reflected beam. We develop an analytical model of this effect, design a realistic setup based on off-the-shelf components to observe it, and simulate the complete experiment, from illumination to detection, with an open-source framework. In the regime where the interference is maximized, the spatial structure of the reflected beam becomes extremely sensitive to the outer medium. This allows us to define an observable more than an order of magnitude more sensitive than amplitude-based measurements, and we expect a comparable improvement in resolution.
\end{abstract}

\maketitle

\section{Introduction}

Surface plasmon resonance (SPR) lies at the heart of plasmonics, and is often considered textbook physics~\cite{raether1988,novotny2006principles}: it is the flagship application of the field, and even a commercial success in molecular detection, from drug discovery to biomarker analysis~\cite{homola2008surface,masson2017surface}. This success rests on a resonant multilayered structure simple enough to be thoroughly understood, easily optimized, and reproducibly manufactured.

These sensors operate by monitoring an observable tied to the resonance~\cite{piliarik2009surface}. SPR detection traditionally relies on measuring changes in the reflection coefficient — a single complex number that fully characterizes how a {\em plane wave} reflects from a multilayered structure. This coefficient exhibits a sharp minimum near the plasmon resonance angle, and monitoring shifts in this minimum forms the basis of conventional SPR sensing. Other observables have been proposed in the past, based on the rapidly varying phase of the reflection coefficient for instance. 

Yet the reflection coefficient, while mathematically simple, emerges from a rich underlying physics. The reflection from a metal film results from the interference of two distinct physical processes : direct Fresnel reflection at the prism-metal interface, and radiative leakage from the excited surface plasmon. Near resonance, these components have nearly equal amplitudes but opposite phases, producing the characteristic reflectance dip through destructive interference. This interferometric nature is well-known theoretically but has been considered merely as an alternative description of the same phenomenon.

However, these two interfering components have fundamentally different spatial characteristics~\cite{tamir71a,simon2007observation}. The Fresnel reflection is local -- the beam reflects exactly where it strikes the interface. In contrast, the plasmon component involves the excitation of a surface wave that propagates along the metal surface and radiates back into the prism. This propagation spatially displaces and broadens the plasmon contribution relative to the direct reflection. For the millimeter-wide beams used in conventional SPR, this spatial difference is negligible — both contributions remain essentially superposed, justifying the single reflection coefficient description. However, when the beam width shrinks to a few tens of microns — typically a dozen times the plasmon propagation length, which radiative leakage reduces to a few microns on the thin films used for SPR — the two contributions begin to spatially separate~\cite{tamir71a,bertoni1973unified}. The destructive interference that normally reduces the overall reflected intensity now becomes localized in space, creating a dip within the reflected beam profile~\cite{kou1989excitation,polles2016leveraging}, which we call the Interferometric Reflectivity Dip (IRD) in the following.

In a regime where the incident beam is angularly wider than in a regular SPR experiment, reflection from the metal film begins to act as a spatial filter. For weakly focused beams, the response of the surface wave amounts to a lateral shift of the reflected beam, which can itself be used to track the resonance~\cite{yin2004large,yin06}; for tightly focused ones it is confined to the near field~\cite{weeber2011near}. In the intermediate regime where the IRD appears, a new family of observables can be envisaged: one can track the variation of the intensity at specific locations, close to where the spatial interference is most pronounced. As this destructive interference is extremely sensitive to any refractive index change, this improves the sensitivity of the device and, potentially, the resolution.

This paper is organized as follows. We first develop an analytical theory of the IRD, describing in simple mathematical terms the interference mechanism outlined above and clarifying the physics at play — in particular the role of the ratio between beam width and plasmon propagation length. We then design a realistic setup based on simple, readily available optical components, and introduce an open-source simulation framework to model the full experiment, from illumination to detection~\cite{tristan_miralles_2026_22282166}. Using these tools, we optimize the setup parameters to maximize the visibility of the effect. Finally, we estimate the performance that can be expected from an IRD-based measurement under realistic conditions, in terms of both sensitivity and resolution, and discuss the noise regimes in which this approach provides an advantage. All codes and parameters are provided so that every result can be reproduced, adapted, or challenged.

\section{Theory of the reflectivity dip}\label{sec:theory}

\begin{figure*}[htbp]
    \centering
    \includegraphics[width=\linewidth]{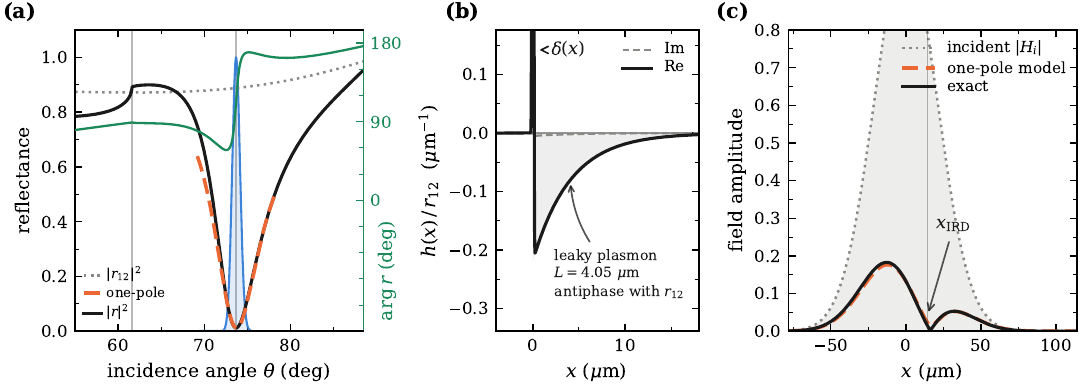}
\caption{Formation of the interferometric reflectivity dip in the Kretschmann
configuration: BK7 prism, 50~nm gold film, water superstrate, $\lambda=632.8$~nm,
$p$ polarization. This simplified structure and the waist are chosen to make the effect clearly visible, and differ from the optimized setup of Section~\ref{sec:optim}. (a) Reflectance versus internal incidence angle. The bare prism--gold interface
$|r_{12}|^{2}$ (dotted) provides a slowly varying background. The vertical lines mark the
critical angle of the prism--water interface and the
resonance. The dashed curve is the one-pole model drawn over the interval in which it is meaningful. The shaded curve is the angular intensity spectrum of a Gaussian beam of waist $w=30$~$\mu$m along the interface. Its angular width is about five times smaller than that of the resonance.
(b) Envelope of the impulse response $h(x)/r_{12}$. The $\delta$ contribution is drawn as a narrow Gaussian and is clipped. The leaky plasmon is essentially in antiphase with the specular reflection; the imaginary part (dashed) is negligible throughout.
(c) Amplitude of the reflected field for $w=30$~$\mu$m, normalized to the peak of the incident beam (dotted); exact calculation (solid) and one-pole model (dashed). The vertical line is $x_{\rm IRD}=14.3$~$\mu$m, as given by Eq.~\eqref{eq:xird}.}
\label{fig:modele}
\end{figure*}

We present here a simplified theory of the Interferometric Reflectivity Dip and compare the results of our simplified theory to bidimensional simulations.  The analysis below follows the approach of Tamir and Bertoni~\cite{tamir71a,kou1989excitation}, who described the reflection of a bounded beam on a leaky-wave structure by an analytic, complex-plane treatment of the reflection coefficient. Their interest was the lateral displacement of the reflected beam; the distortion of its profile
appeared there mainly as a nuisance, since a displacement becomes ill-defined once the beam is strongly reshaped. We focus instead on that distortion, and specifically on the zero their analysis reveals in the reflected profile --- the signature of a destructive interference, and, as we have shown~\cite{polles2016leveraging}, of a marked narrowing of the reflected beam.

We consider a three-layer system consisting of a prism (medium 1, refractive index $n_1$), a thin metal film of thickness $h$ (medium 2, permittivity $\epsilon_m$), and a dielectric sample (medium 3, refractive index $n_3$). For a p-polarized plane wave with transverse wavevector $k_x = n_1 k_0 \sin \theta$ ($\theta$ being the incidence angle and $k_0 = \frac{\omega}{c}$) the total reflection coefficient can be written as:
\begin{equation}
r_{\text{tot}}(k_x) = r_{12} + \frac{t_{12}t_{21}r_{23}\,e^{-2\kappa h}}{1 - r_{21}r_{23}\,e^{-2\kappa h}}
\label{eq:rtot}
\end{equation}
where $r_{ij}$ and $t_{ij}$ are the Fresnel reflection and transmission coefficients at the interface between media $i$ and $j$, and $\kappa = \sqrt{k_x^2 - \epsilon_m k_0^2}$.

This classical expression explicitly separates two distinct physical processes. The first term, $r_{12}$, represents direct Fresnel reflection at the prism-metal interface. This coefficient varies slowly with $k_x$ and provides a nearly constant background reflection. The second term describes plasmon-mediated reflection: light transmitted into the metal layer excites a surface plasmon at the metal-sample interface, which radiates back into the prism. This second term is negligible except close to the resonance, where its amplitude becomes comparable to $r_{12}$ but with opposite phase. This produces a deep minimum in $|r_{\text{tot}}|$, close to zero, at a specific $k_{\min}$ value near the resonance, which can be interpreted as destructive interference between the direct Fresnel reflection and the plasmon contribution.

The surface plasmon excited in the present case is a guided mode at the interface between metal and medium 3 and thus corresponds to a pole of $r_{23}$ for a specific complex value $k_{\text{sp}}^0$. However, $r_{tot}$ does not have any pole for this precise value because $r_{23}$ is present both in the numerator and denominator of the second term of \eqref{eq:rtot}. Instead, the denominator vanishes for a value of $k_x$ that we denote $k_{\text{sp}}$ and which is close enough to the pole of $r_{23}$ so that its modulus can compensate the factor $e^{-2\kappa h}$. The larger the metal film thickness $h$, the closer $k_{\text{sp}}$ is to the complex wavevector of the unperturbed surface plasmon $k_{\text{sp}}^0$. The imaginary part of $k_{sp}$ can be written $1/L$ where $L$ is the field-amplitude decay length, twice the usual intensity propagation length. For our simplified system (BK7 glass / 50~nm Au / water at $\lambda = 632.8$~nm), numerical calculations using PyMoosh~\cite{langevin2024pymoosh} yield $L \approx 4~\mu$m, a value that is typical of gold layers used in SPR sensing~\cite{raether1988}. 

Near $k_{\text{sp}}$, the second term of equation \eqref{eq:rtot} is dominated by the vanishing denominator, so that $r_{tot}$ is well approximated by a single pole above a constant background:
\begin{equation}
r_{\text{tot}}(k_x) \approx r_{12} + \frac{\alpha}{k_x - k_{\text{sp}}}
\label{eq:approx}
\end{equation}
where $\alpha$ is a slowly-varying complex amplitude encoding the strength of the plasmon contribution. Both $\alpha$ and $r_{12}$ are treated as constants over the narrow spectral range of interest.

In most SPR experiments the reflectance minimum is very deep, but it does not reach exactly zero: at resonance the reflection coefficient retains a small residual value $r_{\min}$. Complex plane maps computed with PyMoosh~\cite{langevin2024pymoosh} show that this minimum is located at $k_{\min} \simeq
\text{Re}(k_{\text{sp}})$, meaning that the zero and the pole share the same real part, so that $k_{\text{sp}} = k_{\min} + i/L$. Evaluating
\eqref{eq:approx} at $k_{\min}$ then yields
\begin{equation}
  r_{12} + \frac{\alpha}{k_{\min} - k_{\text{sp}}} = r_{12} + i\alpha L
  = r_{\min},
  \label{eq:condition}
\end{equation}
so that $\alpha \simeq i\,(r_{12} - r_{\min})/L$. The residual
reflection is not a detail here: as shown below, it is what places the
dark fringe inside the reflected beam. The reflection coefficient of the
structure can finally be approximated by
\begin{equation}
  r_{\text{tot}}(k_x) \simeq r_{12}
  + \frac{i\,(r_{12}-r_{\min})/L}{k_x - k_{\text{sp}}}.
  \label{eq:r_final}
\end{equation}

The spectral structure of $r_{\text{tot}}(k_x)$ translates directly into a spatial response. For a beam with incident field amplitude $\tilde{H}_i(k_x)$ in the spectral domain, the reflected field is the inverse Fourier transform of the product of the angular spectrum of the incident field with the reflection coefficient :
\begin{equation}
H_r(x) = \frac{1}{2\pi} \int  r_{\text{tot}}(k_x) \tilde{H}_i(k_x) e^{ik_x x} dk_x,
\end{equation}
which means this Fourier transform can be written as the convolution of the incident beam profile with an impulse response $h(x)$
\begin{equation}
H_r(x) = H_i(x) \otimes h(x),
\label{eq:convolution}
\end{equation}
this impulse response being the inverse Fourier transform of \eqref{eq:r_final}.
The inverse transform of the first term of $r_{tot}$ is approximately a $\delta$ distribution in the spatial domain because over the range of interest where the incidence angular spectrum is nonzero the reflection coefficient indeed appears constant. The second term has a single pole at $k_{\text{sp}}$, and its inverse transform follows from the residue theorem. Since $\text{Im}(k_{\text{sp}}) = 1/L > 0$, the contour closes in the upper half-plane for $x > 0$, yielding 

\begin{align}
    h(x) &= r_{12}\,\delta(x)
          - \frac{r_{12}-r_{\min}}{L}\,e^{ik_{\text{sp}}x}\,\Theta(x)\\
         &= r_{12}\,\delta(x)
          - \frac{r_{12}-r_{\min}}{L}\,e^{ik_{\min}x}\,e^{-x/L}\,\Theta(x)
\end{align}

where $\Theta(x)$ is the Heaviside step function. This second term represents the surface plasmon that propagates along the interface (oscillatory factor $e^{ik_{\min}x}$), decays exponentially with characteristic length $L$, and radiates back into the prism—but only for $x > 0$. The minus sign reflects the destructive interference condition~\eqref{eq:condition}: superposed, the two contributions would cancel up to the residual $r_{\min}$, and it is their spatial separation that makes the cancellation --- the IRD --- observable.

It is actually possible, as shown in the Appendix, to express the convolution product of the simplified impulse response by an incident gaussian beam  defined by $H_i(x) = \exp(-x^2/w^2) e^{ik_{\min} x}$ with an incidence angle $\theta$ defined by $k_{\min} = n k_0 \sin \theta$. The closed form is exact but hardly transparent. When the beam is wide compared to $L$ (see Appendix), however, the reflected beam profile reduces to the incident Gaussian profile multiplied by a ramp,
\begin{equation}
    H_r(x) \simeq r_{\min}\left(1 - \frac{x}{x_{\mathrm{IRD}}}\right) H_i(x),
    \label{eq:ramp}
\end{equation}
where 
\begin{equation}
    x_{\mathrm{IRD}} = \frac{r_{\min}}{r_{12}-r_{\min}}\,\frac{w^2}{2L}.
        \label{eq:xird}
\end{equation}

The reflected beam is thus the incident one, attenuated by the residual reflectivity and multiplied by a factor that vanishes at $x_{\mathrm{IRD}}$: the envelope stays Gaussian, and the whole spatial structure of the IRD is carried by a single zero crossing. The position of that crossing is set by how completely the plasmon cancels the direct reflection at resonance: the deeper the resonance, the smaller $r_{\min}$, and the closer the fringe to the beam centre. It is a finite $r_{\min}$ that pushes it onto the flank, where it is both visible and usable. Since $r_{12}$ and $r_{\min}$ are complex, the linear factor cannot vanish on the real axis, and the IRD is a deep minimum rather than a strict zero. What remains is the destructive interference that distorts the reflected beam by cancelling the right part of the gaussian, even in the absence of a clear zero. We underline that the model contains no adjustable parameter: $L$ follows from the complex pole and $r_{\min}$ from the residual reflection, both computed rather than fitted. Its agreement with the exact simulation is quantitative, as shown in Fig.~\ref{fig:modele}.

Equation~\eqref{eq:xird} gives the typical position for which this destructive interference is the most pronounced and also delimits the regime in which the IRD can be observed. The dark fringe lies within the reflected beam only as long as $x_{\mathrm{IRD}} \lesssim w$, and the expansion above assumes $w \gtrsim L$. Together, these two conditions confine the effect to an
intermediate window
\begin{equation}
    L \;\lesssim\; w \;\lesssim\;
    2\,\frac{r_{12}-r_{\min}}{r_{\min}}\,L .
    \label{eq:regime}
\end{equation}
Within this window, the plasmon tail and the direct reflection separate enough for their destructive interference to produce a localized dip on the flank of the beam. Above it, the dip is pushed beyond the beam envelope: the reflected profile reduces to a weak, attenuated copy of the incident Gaussian and the plasmon response is effectively drowned in it. One then recovers the conventional regime, in which the resonance shows up only through the integrated reflectance and not through any spatial structure. The deeper the resonance, the wider this window: a small $r_{\min}$ is what makes the IRD accessible over a broad range of beam widths.

In the IRD regime, the reflected beam is strongly reshaped: it becomes narrower than the incident beam, as pointed out in our previous work~\cite{polles2016leveraging}. This narrowing is a global signature of the IRD — it can be computed from the beam profiles without locating the fringe itself — and we will use it as such in the following, to find the configurations where the IRD is strongest.

\section{Experimental setup}\label{sec:setup}
 
We now turn to the design of a setup able to observe the IRD and to exploit it for sensing. The previous section gives the requirements: the incident beam width must stay in the regime where the IRD appears, given by Eq.~\ref{eq:regime}. Concretely, the setup must (i) deliver a p-polarized beam with a precisely controlled waist close to the interface, and (ii) image the reflected beam with enough spatial resolution to resolve the dip.

Functionally, such a setup is a coherent imaging system: the first optical system forms an image of the pinhole inside the prism, and the second one relays this image onto the CCD. The SPR multilayer inserted along the way acts as the pupil of this system. Its transmittance is the reflection coefficient $r(\theta)$, which is naturally defined in the space of propagation directions, so that the Fourier plane of the multilayer is a plane of positions: the plane of the waist, which the second optical system images onto the camera. As in any coherent imaging system, the image of a point source is then the Fourier transform of the pupil, here the impulse response $h(x)$ of Eq.~\eqref{eq:convolution}, and the camera records the geometrical image of the pinhole convolved with it. Unlike a physical pupil, which has to be placed in a plane where positions and directions are exchanged, the multilayer acts on directions directly, and can therefore be located anywhere along the beam, including in the image plane itself (see Sec.~\ref{sec:simulation}). In TE polarization, where no plasmon is excited and $r$ is nearly constant over the angular spectrum of the beam, this pupil is flat and the camera sees a faithful, magnified image of the pinhole. In TM polarization, $h(x)$ extends over the plasmon propagation length $L$: as long as the image of the pinhole is much wider than $L$, the impulse response is not resolved and one recovers conventional SPR, while in the window of Eq.~\eqref{eq:regime} it becomes visible, and the IRD appears.

The setup we study can be understood as a coherent imaging system with a filter (the metallic film and the resonance it supports) inserted along the way. The beam waist is imaged onto a camera, and the SPR multilayer acts between the two as a coherent transfer function: each plane wave of the beam is multiplied by the reflection coefficient $r(k)$. Unlike a mask placed in the Fourier plane of a  $4f$ system, this filter acts on the direction of propagation itself and therefore requires no Fourier plane. The multilayer is a non-local optical element: its impulse response extends over the plasmon propagation length $L$. The IRD is simply what this impulse response looks like on the camera, once the beam is narrow enough for $L$ to matter.

To generate a Gaussian beam with an adjustable waist, we start from a well-defined Gaussian beam and magnify it with lenses. The initial beam is produced by a 20\,$\mu$m pinhole illuminated with a helium-neon laser at $\lambda = 632.8$\,nm; measurements show that this yields a Gaussian beam with a waist of 5.69\,$\mu$m (a full width at half maximum of 6.7\,$\mu$m in intensity). A pinhole is only one way to obtain such a beam, and not the most efficient one, since spatial filtering discards most of the source power. Coupling the laser into a single-mode fiber, or directly focusing its TEM$_{00}$ output with a single lens, would deliver the same clean
Gaussian profile with a much higher transmission. This choice matters for the photon budget discussed in Section~\ref{sec:resolution}: any improvement of the source throughput directly pushes back the shot-noise limit. We keep the pinhole here as a conservative, well-characterized baseline.

\begin{figure}[!htbp]
    \centering
    \includegraphics[width=\linewidth]{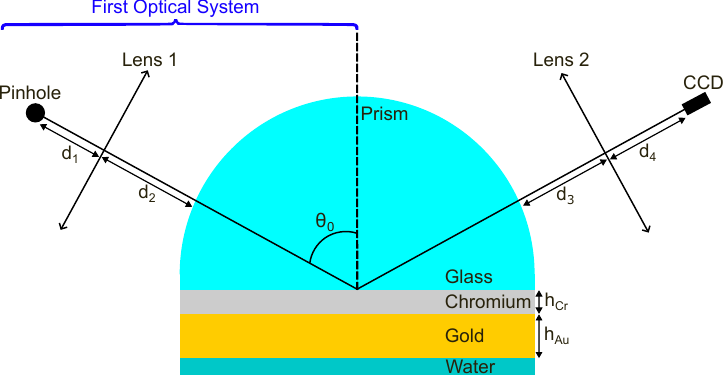}
    \caption{Scheme of the experimental setup considered. The pinhole to first lens distance is $d_1$, the first lens to entrance spherical diopter distance is $\mathrm{d_2}$, the exit spherical diopter to second lens distance is $\mathrm{d_3}$, the second lens to CCD distance is $\mathrm{d_4}$, the central incidence angle of the beam is $\mathrm{\theta_0}$, the chromium layer thickness is $\mathrm{h_{Cr}}$ ($2\,\mathrm{nm}$ here), and the gold layer thickness is $\mathrm{h_{Au}}$.} 
    \label{fig:setup}
\end{figure}

We choose to use a plano-convex lens as the prism itself. The beam then crosses a single spherical diopter, which is simple to take into account in a geometrical description. We consider a standard, nearly hemispherical plano-convex lens. The corresponding parameters are given in the simulation code. We call \emph{first optical system} the ensemble formed by the pinhole, a first lens with a focal length of 30\,mm, a polarizer (not shown on the figure) and the entrance diopter. This system sets the waist at reflection, which is controlled by adjusting the position of the first lens (see Fig.~\ref{fig:setup}). Any standard lens of comparable focal length would do: as shown in section~\ref{sec:simulation}, the image of the beam only needs to lie inside the prism, which leaves ample freedom in the choice and placement of the optics.

The multilayer structure deposited on the flat face of the lens is, from the glass down: a 2\,nm chromium adhesion layer; a gold layer of thickness $h_{\mathrm{Au}}$, in which the surface plasmon is excited — note that $h_{\mathrm{Au}}$ controls the radiative leakage and hence the propagation length $L$, making it one of the three parameters that govern the IRD; and water, standing for the aqueous sample analyzed in commercial SPR instruments.

The chromium layer supports no guided mode of its own and only perturbs the surface plasmon: it adds losses, and shifts the wavevector and the propagation length a little. The pole-zero structure of $r_{\rm tot}$ is
unchanged, so the theory of Section~\ref{sec:theory} applies as it stands, with $L$ and $r_{\min}$ computed for the actual four-layer stack.

The effect is thus controlled by three adjustable parameters: the waist at reflection (set by the first optical system), the central incidence angle $\theta_0$, and the gold thickness $h_{\mathrm{Au}}$. Their optimization is the object of Section~\ref{sec:optim}. Figure~\ref{fig:setup} shows the setup with all parameters.

After reflection, the beam exits through the spherical diopter and a second lens, with a focal length of 35\,mm, images it onto a CCD placed at its image plane. This \emph{second optical system} -- the exit diopter, the second lens, optionally a second polarizer acting as an analyser and the detector -- plays no role in the physics; its only purpose is to match the scale of the reflected beam to that of the detector.
 
The magnification of the second optical system simply has to be large enough for the CCD to resolve the fringe. Our simulation describes the beam with a 2\,$\mu$m step, and the CCD has 26\,$\mu$m pixels, so a magnification of a few tens is sufficient. This is a matter of matching the detector at hand, not a constraint on the physics, and it leaves the choice of the second lens essentially free.

The setup can also be understood from the point of view of Fourier optics, as a coherent imaging system: the first optical system forms an image of the pinhole inside the prism, and the second one relays this image onto the CCD. The SPR multilayer inserted along the way acts as an angular filter, whose coherent transfer function is the reflection coefficient $r(\theta)$. Since $r$ depends on the direction of propagation, its Fourier transform, the impulse response $h(x)$ of Eq.~\eqref{eq:convolution}, is a function of position. As in any coherent imaging system, the image of a point source is the Fourier transform of the coherent transfer function: the camera, conjugate to the plane of the waist, records the geometrical image of the pinhole convolved with $h(x)$.
Because it acts on directions of propagation, the multilayer does not need to lie in any particular plane, which is fortunate: tilted by $\theta_0$ with respect to the beam, it lies in none of them.

\section{Hybrid simulation method}\label{sec:simulation}

While Fourier optics could in principle describe exactly how the beam is reshaped by the lenses and by the reflection, doing so would be computationally too costly. We instead assume that the lenses and diopters form perfect geometrical images, and reserve Fourier optics for the one step where it is unavoidable: the reflection on the gold film, treated as a transfer function in the Fourier plane. Indeed, each plane wave of the beam spectrum experiences its own reflection coefficient, and that coefficient varies rapidly with the angle --- which is precisely the origin of the IRD. Our method is therefore hybrid, combining geometrical and Fourier optics. The full simulation is available as an open-source notebook~\cite{tristan_miralles_2026_22282166}, so that every result below can be reproduced or adapted.

The simulation starts from the beam at the pinhole. The incident light is p-polarized, so that the magnetic field lies along $y$. We follow this single component $H_y$ throughout, and denote it $H$ for brevity. Since we consider a beam centered both spatially and spectrally, its field on the plane orthogonal to the optical axis, assuming unit amplitude, reduces to
\begin{equation}
  H(x, y) = e^{-\left(\frac{x}{w}\right)^2
               -\left(\frac{y}{w}\right)^2},
  \label{eq:beam}
\end{equation}
where $w$ is the initial waist --- the same convention as for the incident profile $H_i(x)$ of Section~\ref{sec:theory}, so that the intensity full width at half maximum is $w\sqrt{2\ln 2}$. The code also accepts spatial and angular offsets of the beam for more general studies. The field is sampled on a square window of size $d = 400\,\mu$m with a $2\,\mu$m step, i.e. $2N+1 = 201$ points per axis, and we switch between the spatial and spectral domains through the Discrete Fourier Transform, with spectral sampling interval $2\pi/d$. These values are a trade-off: the window must be large enough to avoid artifacts from the implicit periodization, and the sampling fine enough to resolve the beam.

For a given central incidence angle $\theta_0$ and a target waist at reflection, the simulation builds a realistic first optical system providing the corresponding magnification. The focal length of the first lens is fixed beforehand (30\,mm here); the adjustable quantities are the pinhole--lens and lens--diopter distances, which the simulation scans over realistic ranges until the target magnification is met.

Consider now what happens inside the prism: the beam propagates in a homogeneous medium of index $n$, reflects on the multilayer, and propagates again towards the exit diopter. In the Fourier domain, both steps are multiplications. Propagation over a distance $l$ multiplies each plane wave by a phase factor $e^{ik_z l}$, and reflection multiplies it by the reflection coefficient $r$ of the stack. Writing a propagation over $l_1$, a reflection, and a propagation over $l_2$,
\begin{align}
  \widetilde{H}(k_x, k_y; l_1 + l_2)
  &= \widetilde{H}(k_x, k_y; 0)\,
     e^{i k_z l_1}\, r(k_x, k_y)\, e^{i k_z l_2} \nonumber \\
  &= \widetilde{H}(k_x, k_y; 0)\,
     e^{i k_z (l_1 + l_2)}\, r(k_x, k_y),
  \label{eq:commute}
\end{align}
these two operations commute, and the total phase depends only on $l_1 + l_2$ -- not on where along the path the reflection occurs. The beam waist therefore need not be located on the gold surface: it can be placed anywhere inside the prism, and the beam footprint on the gold may be larger than the waist without changing the result, since only the angular spectrum matters. This relaxes the constraints on the experimental setup as well as on the simulation. What matters instead is that the camera be conjugated to the plane of the waist.

In the simulation we apply the reflection where the first optical system forms the image of the pinhole, since this is where the beam is narrowest and best sampled by our grid. The propagation phase then never has to be computed at all: the second optical system re-images that same plane onto the CCD, so the field we compute at the reflection is the field the detector sees, magnified by the second optical system.

The reflection coefficient of the stack is computed with PyMoosh~\cite{langevin2024pymoosh} for each plane wave of the spectrum. The materials are described as follows: for BK7, PyMoosh's experimental dispersion model, the RefractiveIndex.info data not covering our wavelength; for chromium and gold, the Brendel-Bormann model from the same database~\cite{polyanskiy2024,rakic1998optical}; for water, a constant index $n_{\mathrm{H_2O}} = 1.33$, so that the sensitivity can later be evaluated by varying it directly.

Strictly speaking, the plane waves of a three-dimensional beam are not all $p$-polarized: for a plane wave with $k_y \neq 0$ the plane of incidence is slightly rotated, so that a field imposed along $y$ -- which a linear polarizer does in practice -- acquires a small $s$ component. This polarization mixing upon beam reflection is a classical and well-documented effect, and it is what makes circularly polarized beams undergo a transverse shift upon reflection~\cite{bliokh2013goos}. The linearly polarized beam considered here carries no such asymmetry, and the mixing is small enough for the whole spectrum to be treated as $p$-polarized; the cross-polarized component can in any case be entirely removed by an analyser aligned with $y$ in front of the detector.

Each plane wave can then be treated as a two-dimensional problem, which is all PyMoosh needs: the multilayer is laterally invariant at the scale of the beam, so every plane wave sees the same stack and reflects within its own plane of incidence, independently of the others. The incidence angle of a plane wave of wave vector $(k_x, k_y, k_z)$ is 
\begin{equation}
  \theta = \arcsin\!\left(
    \frac{\sqrt{(k_x \cos\theta_0 + k_z \sin\theta_0)^2 + k_y^2}}
         {n k_0}
  \right),
  \label{eq:angle}
\end{equation}
where $\theta_0$ is the central incidence of the beam, $n$ the index of the prism, and $k_0 = 2\pi/\lambda$. Multiplying the spectral field by the corresponding reflection coefficients and transforming back to the spatial domain yields the reflected beam.

At this point the simulation is complete: from the beam leaving the pinhole to the light intensity collected by each pixel of the CCD, every step of the experiment is accounted for. We can therefore predict what an experimentalist would actually measure, for any choice of the parameters. The theory of Section~\ref{sec:theory} tells us that the IRD only appears within a window of beam widths, but it does not tell us where, inside that window, the effect is most pronounced, nor what should be measured. Moreover, in Section~\ref{sec:theory}, $k_x$ is the wavevector along the interface and $w$ the waist in that direction, whereas the waist set by the optical systems is measured in the plane transverse to the beam. Close to the axis, a transverse wavevector $k_x$ corresponds to $k_x\cos\theta_0$ along the interface, so that a transverse waist $w$ corresponds to a waist $w/\cos\theta_0$ along the interface, about $3.5$ times larger in our configuration. It is this projected waist that must be compared to $L$ in Eqs.~\eqref{eq:xird} and \eqref{eq:regime}, and the theory remains a guide rather than a design tool. We therefore address both questions with the full simulation: we first search the parameter space for the configuration that maximizes the effect, and then build, from the intensity distribution on the detector, an observable suited to sensing.

\section{Maximizing the IRD and building a sensing observable}\label{sec:optim}

As seen above, three parameters govern the IRD: the beam waist at reflection, the central incidence angle $\theta_0$, and the gold thickness $h_{\mathrm{Au}}$, which controls the plasmon leakage and hence the propagation length $L$. Before defining any observable, we first look for the configuration in which the IRD is the strongest. To do so, we scan this three-dimensional parameter
space with the full simulation: the waist from the initial 5.69\,$\mu$m up to fifty times this value, the incidence around the resonance angle of the structure, and the gold thickness from 20\,nm --- a practical lower bound to preserve layer quality --- to 55\,nm, a thickness beyond which the resonance becomes difficult to excite.

\subsection{Maximizing the IRD}

At this stage the figure of merit we rely on is the relative enlargement along $x$~\cite{polles2016leveraging}: the second centered moment of the reflected intensity, divided by that of the incident one,
\begin{equation}
  \Xi_x =
  \frac{\displaystyle
        \frac{\iint (x-\delta_x)^2 |H_r|^2\,\mathrm{d}x\,\mathrm{d}y}
             {\iint |H_r|^2\,\mathrm{d}x\,\mathrm{d}y}}
       {\displaystyle
        \frac{\iint x^2 |H_i|^2\,\mathrm{d}x\,\mathrm{d}y}
             {\iint |H_i|^2\,\mathrm{d}x\,\mathrm{d}y}},
  \label{eq:Xi}
\end{equation}
where $\delta_x$ is the shift of the beam barycenter upon reflection. This is simply the squared width of the beam after reflection compared to its width before, and $\Xi_x < 1$ therefore signals a narrowing of the reflected beam -- the global signature of the IRD, as explained in Section~\ref{sec:theory}. The connection with the fringe is direct: the second moment weights the intensity by $x^2$, so removing light on the flank of the beam, where $x$ is large, lowers it appreciably. We stress that $\Xi_x$ is a numerical diagnostic, not a measurable quantity: it requires the knowledge of the incident profile $|H_i|^2$, which is not accessible in an experiment. Its role here is only to locate the regime where the IRD is strongest; the measurable observable is built in the next subsection. The enlargement along $y$ is also computed, and remains equal to one throughout the scan, confirming that the effect only develops along the plasmon propagation direction.

The optimum, $\Xi_x \simeq 0.75$, is reached for a target waist of 46.09\,$\mu$m, an incidence of $\theta_0 = 73.38^\circ$, and $h_{\mathrm{Au}} = 47$\,nm. The realistic first optical system built by the simulation achieves a magnification of 8.12, i.e.\ a waist of 46.19\,$\mu$m --- close enough to the target, given how flat the minimum is, to leave $\Xi_x$ unchanged. The scan also finds slightly deeper minima of $\Xi_x$ for nearby gold thicknesses, but they correspond to very wide beams and to configurations where almost no light is reflected, which are of no practical interest.

Having found the optimum, we now look at how $\Xi_x$ behaves around it. Varying each parameter in turn tells us two things: what has to be controlled in an experiment, and how steep the response to the index really is.

\begin{figure}[!htbp]
    \centering
    \includegraphics[width=\linewidth]{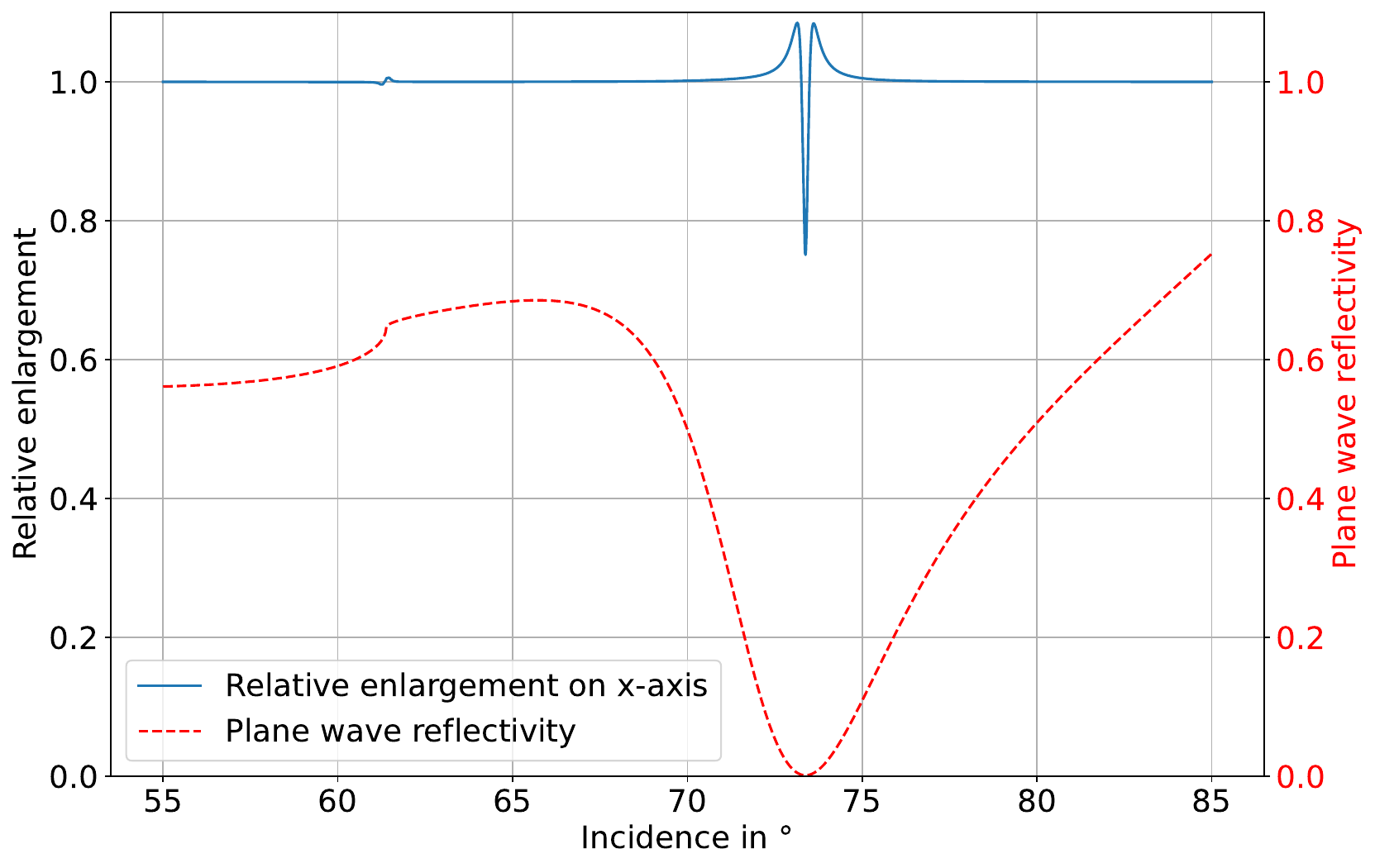}
    \caption{Relative enlargement of the reflected beam along $x$ (blue, left axis) and plane-wave reflectivity of the structure (red, right axis), as a function of the central incidence angle, at the optimal waist and gold thickness.}
\label{fig:scan_inc}
\end{figure}

\begin{figure}[!htbp]
    \centering
    \includegraphics[width=\linewidth]{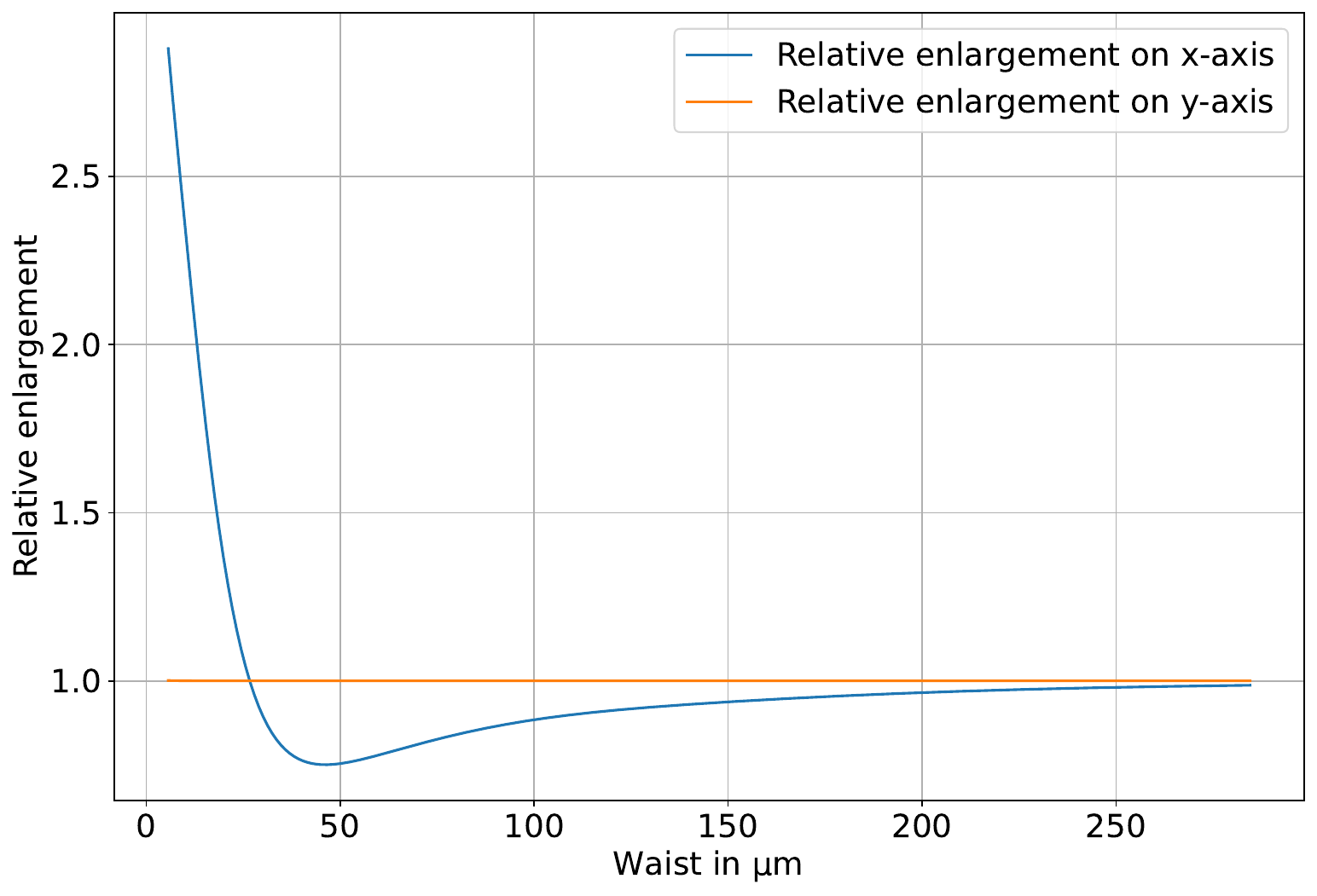}
    \caption{Relative enlargement of the reflected beam as a function of the beam waist at reflection, at the optimal incidence and gold thickness. The narrowing along $x$ (blue) persists over a broad range of waists, in agreement with the window predicted by Eq.~\eqref{eq:regime}.}
    \label{fig:scan_w}
\end{figure}

\begin{figure*}[!htbp]
    \centering
  \includegraphics[width=0.45\textwidth]{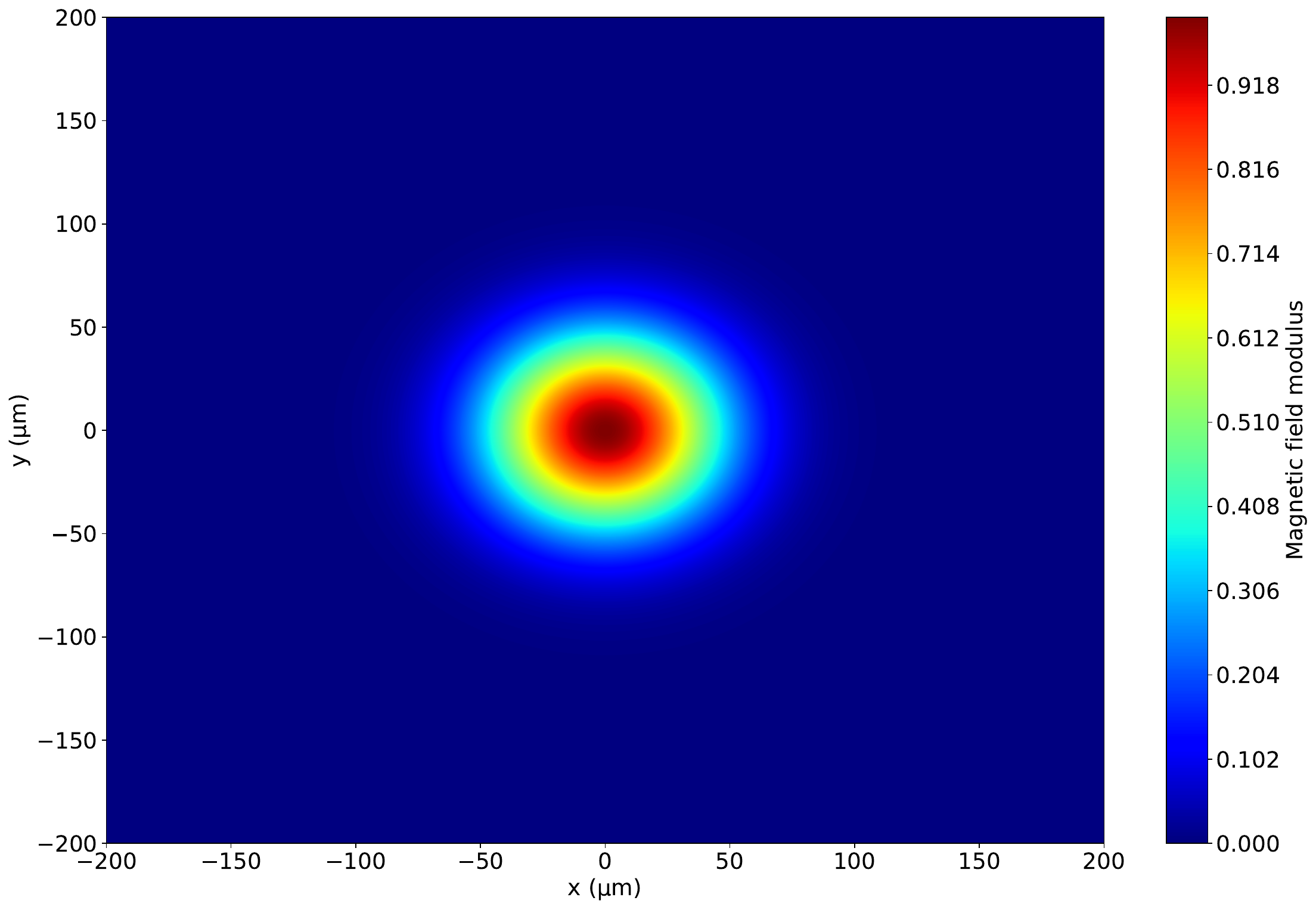}
  \includegraphics[width=0.45\textwidth]{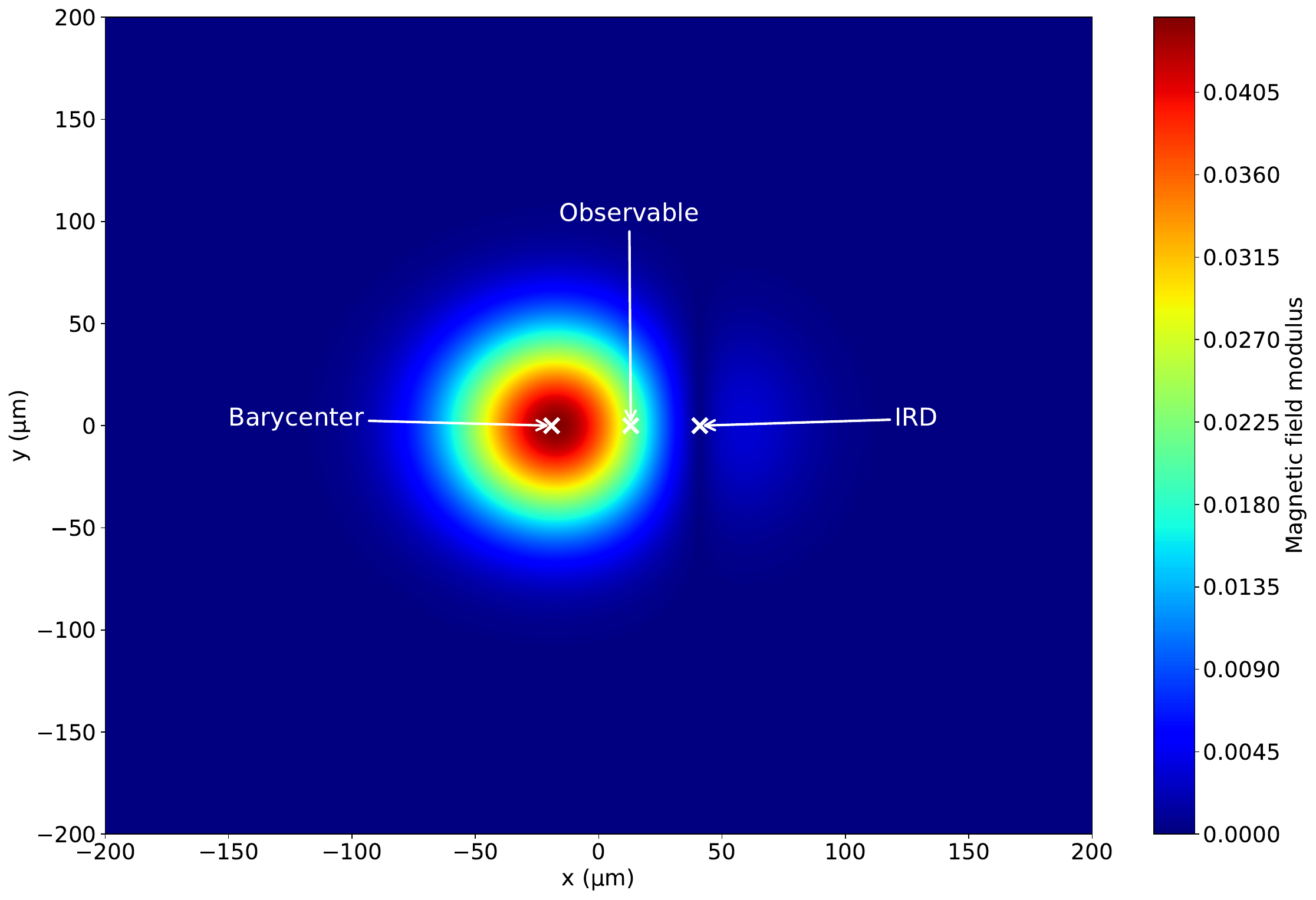}
\caption{Magnetic field modulus in the $(x,y)$ plane, before (left) and after (right) reflection on the multilayer, in the optimal configuration. The scale is that of the beam at the reflection, in the plane transverse to the beam, before the magnification of the second optical system, and all fields are normalized to the maximum amplitude of the incident beam. The IRD appears as a dark fringe on the $x$ side of the reflected beam, whose barycenter is displaced laterally by 19\,$\mu$m to the left. The reflected field peaks at 4\% of the incident amplitude, so the total reflected intensity is of the order of $2\times 10^{-3}$ of the incident one.}
  \label{fig:beam_deformation}
\end{figure*}

The incidence angle is by far the most critical. The narrowing survives only over a very narrow angular range --- far narrower than the resonance itself (Fig.~\ref{fig:scan_inc}) --- since the effect requires the plasmon to be properly excited. An experiment will have to scan the angle slowly and hold it steady.  The waist, in contrast, is forgiving: any value between roughly 35 and 70\,$\mu$m produces a marked IRD (Fig.~\ref{fig:scan_w}), an interval that is essentially the window predicted in Section~\ref{sec:theory}, where the fringe is separated enough from the direct reflection to be visible yet still well inside the beam envelope. The gold thickness is not critical either; changing it mostly shifts the optimal angle. Note that the beam itself spans about $0.2^\circ$ in incidence — narrower than the resonance, but comparable to the angular range over which the IRD develops. We underline that even at the optimal angle, most of its plane-wave components are not at the reflectivity minimum, which is why the reflected intensity does not vanish, as shown in Fig.~\ref{fig:beam_deformation}.

\begin{figure}[!htbp]
    \centering
    \includegraphics[width=\linewidth]{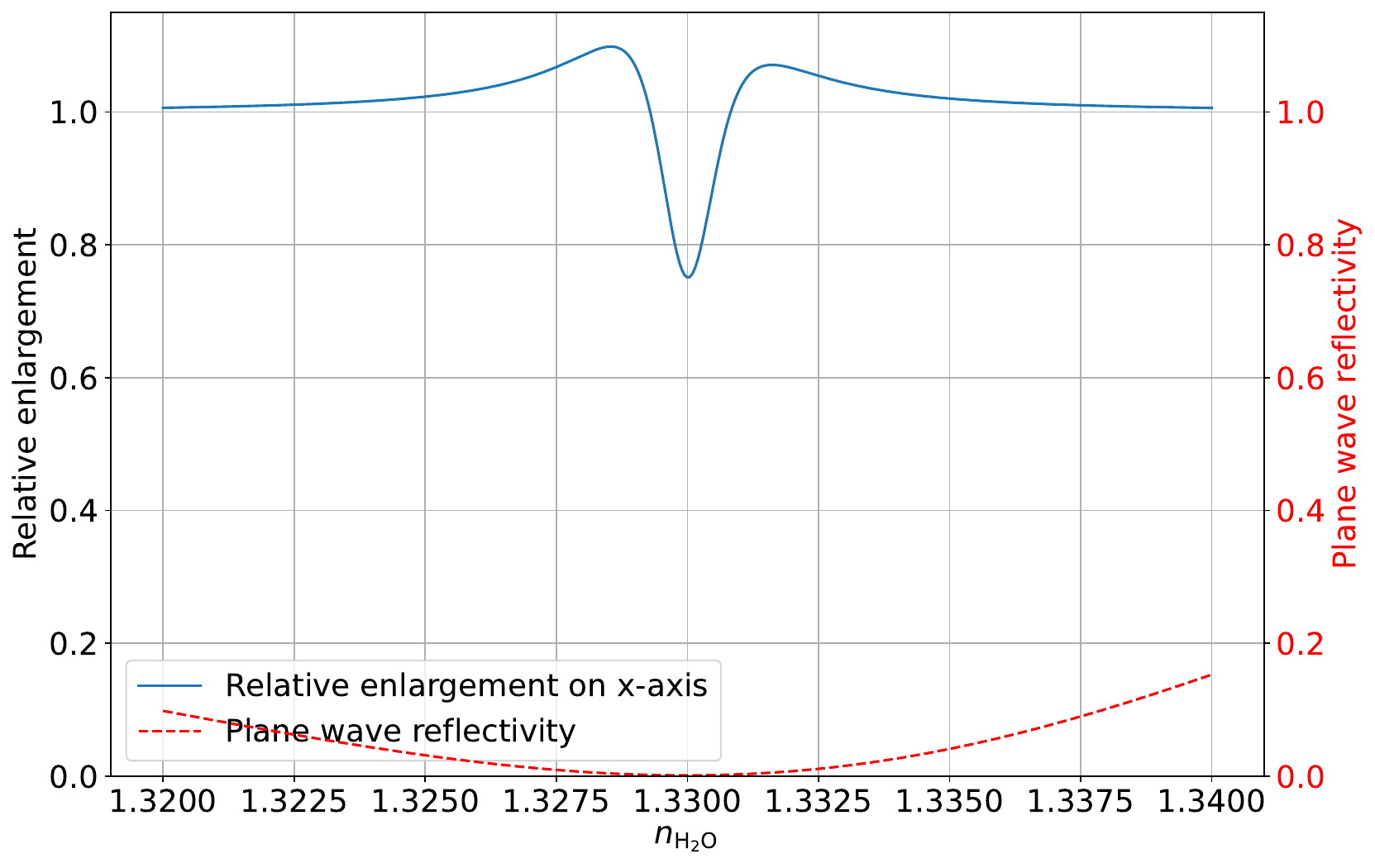}
    \caption{Relative enlargement along $x$ (blue, left axis) and plane-wave reflectivity (red, right axis) as a function of the refractive index of the outer medium, at the optimal waist, incidence and gold thickness. The IRD responds over a range of a few $10^{-3}$ in index, centered on the working point where the reflectivity is at its minimum and nearly flat.}
    \label{fig:scan_n}
\end{figure}

Figure~\ref{fig:scan_n} shows $\Xi_x$ against the index of the outer medium: the IRD develops over a few $10^{-3}$ around the working point, where the reflectance is at its minimum and barely responds at all. This is the mirror image of the angular dependence — both express how finely the interference condition is balanced — and it is where the sensitivity gain reported below comes from. It also fixes the range over which an IRD sensor can operate: a steep response is necessarily a narrow one. 

We underline that we vary the index of the whole outer medium, which is the conventional way of characterizing a sensor. A more realistic model would require knowing what is being detected: a functionalization layer of antibodies or aptamers, then the analyte bound to it, with thicknesses and indices that depend entirely on the assay and the chemistry involved. Simulating such a stack would be straightforward with the tools we provide --- it is a matter of adding layers --- but choosing them meaningfully requires knowing the intended application, which is beyond the scope of this paper. A bulk sensitivity, by contrast, is chemistry-independent, comparable across instruments, and gives a fair idea of what the effect is worth.

Figure~\ref{fig:beam_deformation} shows the beam before and after reflection in this optimal configuration. The narrowing appears as a dark fringe on the right ($x>0$) side of the reflected beam, where Eq.~\eqref{eq:xird} places it. The barycenter of the beam is displaced by about $-19\,\mu$m --- that is, backwards. This deserves a comment. A strongly excited leaky mode is usually expected to shift the beam forward, in the direction the plasmon propagates. Artmann's formula gives the shift of a wide beam as the derivative of the phase of $r$~\cite{felbacq03,moreau08,polles2016leveraging}, and that phase varies rapidly near resonance, so the expected displacement is large. We observe the opposite, and the reason is the fringe itself. The destructive interference removes light precisely on the side towards which the plasmon propagates, and the barycenter recedes accordingly. The negative shift is therefore not a sign that the plasmon is weakly excited: it is the signature of its excitation --- another illustration of how misleading a global, asymptotic description can be when the beam is narrow.

\subsection{From the IRD to a measurable observable}

We work in amplitude-modulation mode: the setup is kept fixed, and we look for the point on the detector where the collected intensity varies
most when the refractive index of the outer medium changes. Scanning the reflected beam, the strongest response is found 28\,$\mu$m before the
fringe, towards the beam centre --- that is, 33\,$\mu$m from the barycenter of the reflected beam, for a waist of 46.19\,$\mu$m. This
places the observable a little past halfway between the beam centre and the fringe, on the rising flank of the dip and not at its bottom (Fig.~\ref{fig:beam_deformation}).

This is where one expects it to be. A change $\delta n$ modifies the residual reflectivity $r_{\min}$ and thus the position of the fringe through Eq.~\eqref{eq:xird}, while detuning the resonance condition that produces it: the dark fringe both moves and fades. A structure that moves is best detected where the intensity profile is steepest, which is the flank. The bottom of the dip is in fact the worst choice: the intensity is stationary there, and the few photons collected make the measurement noisiest --- a point we return to in
Section~\ref{sec:resolution}. What matters is to sit on the flank, not at one precise point: the response varies little within some 10\,$\mu$m either side of the optimum. Finding a sensitive position is therefore easy, and this tolerance is what makes the measurement realistic.

The location of this optimum also supports the model of Section~\ref{sec:theory}. For a Gaussian beam alone, the steepest slope would lie at half the waist from the centre, i.e. 23\,$\mu$m. The reflected profile is not Gaussian, however, but the incident one multiplied by the ramp of Eq.~\eqref{eq:ramp}, which shifts the steepest slope towards the fringe. The optimum found numerically, at 33\,$\mu$m, lies on that side, as expected.

We therefore take as our observable the intensity collected on the line of pixels located there, summed along $y$: a whole row rather than a single pixel, which averages out part of the pixel noise.

\subsection{Sensitivity gain}

\begin{figure}[!htbp]
    \centering
    \includegraphics[width=\linewidth]{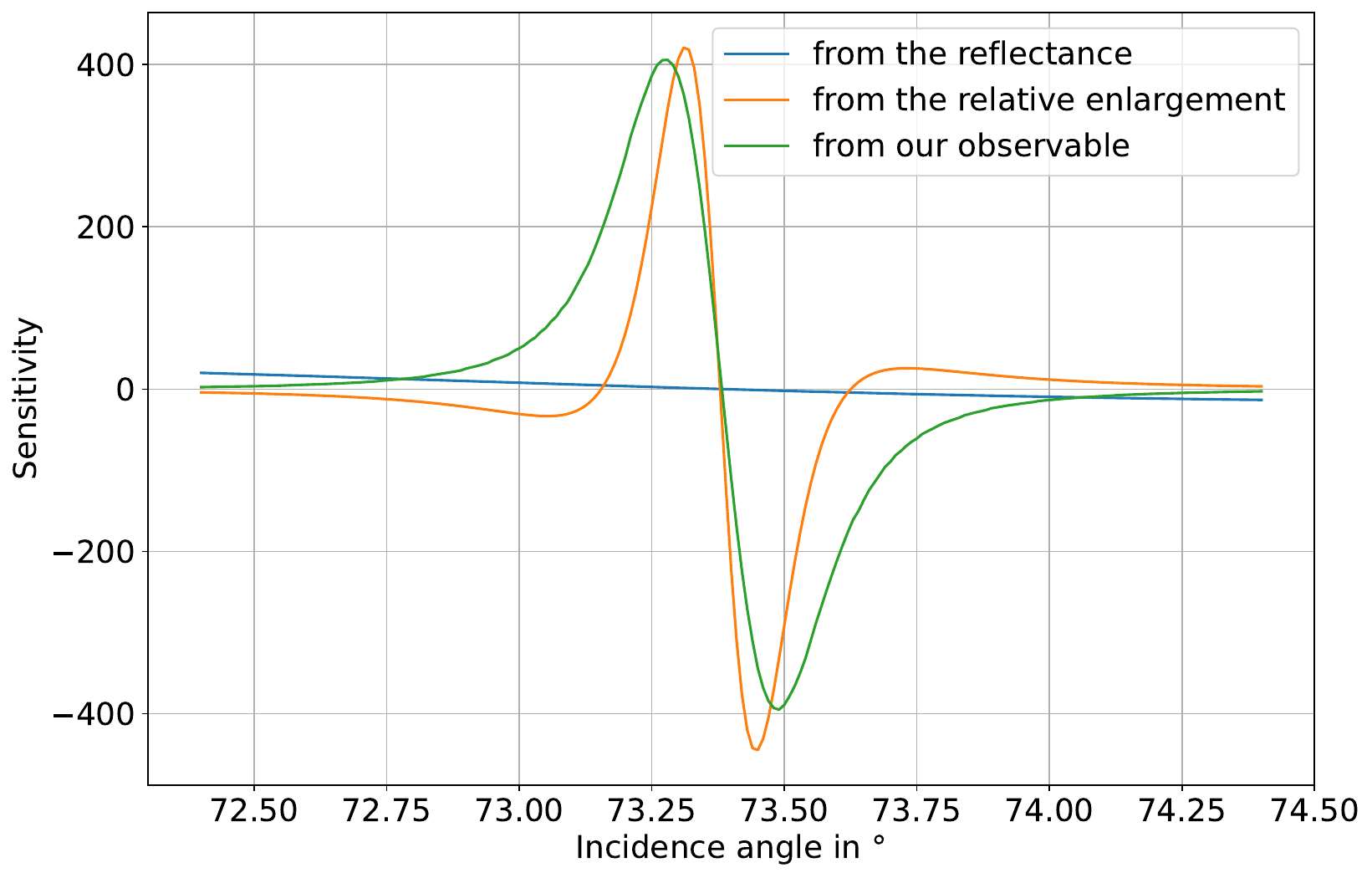}
    \caption{Sensitivity of three observables, determined with an index variation under the prism $\delta n_{\mathrm{H_2O}}=10^{-5}$, as a function of the central incidence angle: the conventional reflectance, the relative enlargement $\Xi_x$, and the intensity on the line of pixels selected on the flank of the fringe. The intensity observable tracks the enlargement closely and peaks an order of magnitude above the reflectance.}
    \label{fig:sensitivity}
\end{figure}

Figure~\ref{fig:sensitivity} compares, as a function of $\theta_0$, the sensitivity of three observables: the conventional reflectance, the relative enlargement $\Xi_x$ --- which is not measurable, but tracks the IRD globally --- and the intensity on our line of pixels. All three
change sign across the optimum: on either side of it the fringe moves in the opposite direction, so the observable rises where it used to fall. What matters here is the magnitude of the response. The intensity observable and the enlargement follow each other closely, reaching comparable extrema at nearly the same angles, which confirms that our measurable observable is a faithful proxy for the effect. The sensitivity linked to the reflectance, on the same scale, is almost flat.

On the structure optimized for the IRD, the largest sensitivities are approximately 400\,RIU$^{-1}$ for the IRD intensity and 420\,RIU$^{-1}$ for the enlargement, against 20\,RIU$^{-1}$ for the reflectance. These are computed with $\delta n = 10^{-5}$, small enough for the values to be stable: they change by less than a few percent when $\delta n$ is varied between $10^{-4}$ and $10^{-5}$, while a step of $10^{-3}$ already averages the response and underestimates the sensitivity by some 15\%. The reflectance figure is not a fair comparison, however, since the gold thickness was chosen to maximize the IRD and not the reflectance sensitivity. Re-optimizing it for the reflectance observable itself brings the latter to about 30\,RIU$^{-1}$, so that the IRD observable remains ahead by an order of magnitude.

The re-optimization is necessary because the two observables push the operating point in opposite directions. The IRD requires a substantial part of the surface plasmon to leak back into the prism: this pushes the fringe away from the beam centre, and leaves more light overall, which helps with noise (Section~\ref{sec:resolution}). A conventional reflectance measurement wants the opposite, the deepest possible dip, since this is what steepens the slope on the side of the resonance.

\section{Noise regimes and resolution}
\label{sec:resolution}

A sensitivity gain does not directly give a resolution gain: the resolution is $\sigma_n = \sigma_Y/S$, and it depends on the noise $\sigma_Y$ as much as on the sensitivity $S$. We have tried to estimate what our order-of-magnitude gain in sensitivity may actually be worth in terms of resolution.

The first concern is that our observable sits on a dark region of the beam, where few photons are collected. At the operating point it receives a fraction $f \simeq 2.4\times10^{-5}$ of the light reaching the sensor. Its relative shot noise, $1/\sqrt{fN}$, falls below a technical noise floor $\sigma$ as soon as 
\begin{equation}
  N \;\gtrsim\; \frac{1}{f\,\sigma^{2}},
  \label{eq:budget}
\end{equation}
which amounts to 13\,nJ of collected light for $\sigma = 10^{-3}$, and 0.13\,mJ for $\sigma = 10^{-5}$ --- microseconds to a fraction of a second of a milliwatt laser. Photon noise thus seems easy to push below
the technical floor, all the more so as a dark pixel is the last one to saturate the sensor. This is also one more reason to work on the flank rather than at the bottom of the dip, where $f$ is several orders of magnitude smaller.

Beyond shot noise, the resolution is set by technical noise, which only a real setup can characterize. If the noise floor is absolute, the sensitivity gain carries over directly to the resolution. If it is multiplicative, as for laser power fluctuations, or comes from beam pointing, the fact that the observable is read on an image offers a handle a single-photodiode measurement lacks: the selected line can be normalized by the total intensity of the beam, and the fringe located relative to the barycenter of the beam, both recorded on the same sensor at the same instant. How well this works is a question for experiment.

Finally, this comparison concerns amplitude modulation only. Commercial instruments interrogate the resonance angularly or spectrally and reach much finer resolutions~\cite{piliarik2009surface}. Phase-sensitive schemes are more sensitive still~\cite{huang2012phase}, but require interferometric or polarization-modulation setups that have hindered their deployment. The IRD retrieves part of this phase sensitivity, since the fringe results from the interference between the direct reflection and the plasmon contribution, with a simple camera and no reference arm.

\section{Conclusion}

This work started from a simple observation: the two contributions that interfere in SPR reflection — the direct Fresnel reflection and the plasmon radiation — do not live at the same place on the interface. When the beam waist is some ten times the plasmon propagation length, their destructive interference becomes visible as a localized dark fringe in the reflected beam, the Interferometric Reflectivity Dip. We have given a simple analytical model of this fringe, designed a realistic setup to observe it with off-the-shelf components, and simulated the whole experiment with an open-source framework.

The IRD is not only observable: it can be useful. The fringe is sensitive to the slightest change of the outer medium, so the intensity on its flank reacts an order of magnitude more strongly than the overall reflectance. Whether this translates into as large a gain in resolution depends on what limits the instrument. Against an absolute noise floor it does, directly. Against source fluctuations, the advantage rests on the observable being read on an image: it can be normalized by the beam itself, measured on the same sensor at the same instant. Only an actual experiment can settle how well that works. We hope the setup and the code provided here will make this experiment easy to attempt, and we anticipate its results with great interest.

We have explored one mode of operation among several possible ones. Amplitude modulation at a fixed angle was the simplest to simulate and to implement, but it is probably not the last word. What first drew our attention to the effect is how much narrower the angular range of the narrowing is than the resonance itself (Fig.~\ref{fig:scan_inc}) --- a sharpness of the same nature as the sensitivity gain reported above, since both measure how finely balanced the same interference condition is, read once in angle and once in space. An angular or spectral interrogation of the IRD therefore seems well worth investigating. More generally, our work shows that the light sent onto an SPR structure is a design parameter in its own right: a Gaussian beam of well-chosen waist and incidence is enough to turn the reflection into a spatially structured and highly sensitive signal. This suggests that more elaborately shaped beams could exploit the reflection on a resonant structure more efficiently still.

Beyond SPR sensing, the physics reported here is general~\cite{bertoni1973unified}. The IRD arises whenever the directly reflected field interferes with a leaky wave of finite spatial extent. Nothing in the mechanism is specific to surface plasmons, and other leaky guided modes should produce it as well~\cite{tamir86,moreau08,krayzel10}. The zero resulting from this destructive interference has been noticed in the past, but seen as a nuisance rather than as a resource~\cite{tamir71a}. This regime has probably been overlooked because it is hidden twice over: by the plane-wave description, which knows nothing of space, and by the asymptotic treatment usual in non-specular optics~\cite{tamir86,yin2004large,bliokh2013goos}, which focuses on the lateral shift of the beam~\cite{yin06,weeber2011near} and draws attention away from its distortion. Both simplifications are made almost everywhere. It seems worth asking what else they may have been hiding.

\section*{Appendix}
    
Here we compute the convolution 
 $$H_r(x)  \simeq  H_i \otimes h$$
 where 
 $$h(x)  = r_{12} \delta(x) - \frac{r_{12}-r_{\min}}{L} e^{i k_{\min} x} e^{-\frac{x}{L}} \Theta(x)$$
with 
$H_i(x) = e^{-\frac{x^2}{w^2}} e^{i k_{\min} x}$ and $\Theta$ is the Heaviside function. 
\begin{widetext}
This gives 
\begin{align}
    H_r(x)  &= r_{12} \, H_i(x) - \frac{r_{12}-r_{\min}}{L} \int_0^\infty  e^{-\frac{(x-u)^2}{w^2}} e^{i k_{\min} (x-u)} e^{i k_{\min} u} e^{-\frac{u}{L}} \, du \\
    & = r_{12} H_i(x) -\frac{r_{12}-r_{\min}}{L}  e^{i k_{\min} x} \int_0^\infty e^{-\frac{(u-x)^2}{w^2}} 
    e^{-\frac{u}{L}} \, du\\
    & =  e^{i k_{\min} x} \left[r_{12}\, e^{-\frac{x^2}{w^2}} - \frac{r_{12}-r_{\min}}{L} \int_0^\infty e^{-\left(\frac{u - x + \frac{w^2}{2L}}{w}\right)^2} e^{\frac{w^2}{4L^2}-\frac{x}{L}} \,du
    \right]\\
    &= e^{i k_{\min} x} \left[r_{12} \,e^{-\frac{x^2}{w^2}} - \frac{r_{12}-r_{\min}}{L} \frac{w \sqrt{\pi}}{2}\, e^{\frac{w^2}{4L^2}-\frac{x}{L}} \, \erfc \left(\frac{w}{2L} - \frac{x}{w}\right) \right]
\end{align}
\end{widetext}

We define $z = \frac{w}{2L} - \frac{x}{w}$ so that $\frac{w^2}{4L^2}-\frac{x}{L} = z^2 - \frac{x^2}{w^2}$ so that the expression becomes

\begin{align}
    H_r(x) &= H_i(x) \left(r_{12} - (r_{12}-r_{\min}) \frac{w \sqrt{\pi}}{2L}\, e^{z^2} \, \erfc(z)\right),
\end{align}
where the Faddeeva function taken in $iz$, $e^{z^2} \, \erfc(z)$, can be recognized. When $z$ is large, which is the case here since $w \gg L$ and the ratio $\frac{x}{w}$ is typically lower than 1, the function is equivalent to $\frac{1}{z\sqrt{\pi}}$ so that 
\begin{align}
    H_r(x) &\simeq H_i(x) \left(r_{12} - (r_{12}-r_{\min}) \frac{w}{2Lz}\right)\\
    &\simeq H_i(x)  \left(r_{12} - \frac{r_{12}-r_{\min}} {1 - 2Lx/w^2}\right).
\end{align}

Finally, since $L\ll w$ and $x <w$, $2Lx/w^2 \ll 1$ in the region of interest, so that this last expression can be expanded to first order to yield
\begin{equation}
        H_r(x) \simeq r_{\min}\left(1 - \frac{x}{x_{\mathrm{IRD}}}\right) H_i(x),
\end{equation}
with
\begin{equation}
    x_{\mathrm{IRD}} = \frac{r_{\min}}{r_{12}-r_{\min}}\,\frac{w^2}{2L}.
\end{equation}

\begin{acknowledgments}
This work was supported by the Agence Nationale de la Recherche
(ANR-23-CE09-0014, SWAG-P; ANR-24-CE47-1304, eQUANDIS).
\end{acknowledgments}

\section*{Data availability}
The code and data underlying the results presented in this paper are
available in Ref.~\cite{tristan_miralles_2026_22282166}.


\bibliography{article}

\end{document}